\documentclass[conference]{IEEEtran}
\IEEEoverridecommandlockouts
\usepackage{cite}
\usepackage{amsmath,amssymb,amsfonts}
\usepackage{algorithmic}
\usepackage{graphicx}
\usepackage{textcomp}
\usepackage{xcolor}
\def\BibTeX{{\rm B\kern-.05em{\sc i\kern-.025em b}\kern-.08em
    T\kern-.1667em\lower.7ex\hbox{E}\kern-.125emX}}
\begin{document}

\title{Evaluating Pedagogical Incentives in Undergraduate Computing: A Mixed Methods Approach Using Learning Analytics  \\
\thanks{This project received funding from UCL ChangeMakers.}
}

\author{\IEEEauthorblockN{Laura J. Johnston}
\IEEEauthorblockA{\textit{Department of Statistical Science}\\
\textit{University College London}\\
London, UK \\
laura.johnston.22@ucl.ac.uk}
\and
\IEEEauthorblockN{Takoua Jendoubi}
\IEEEauthorblockA{\textit{Department of Statistical Science}\\
\textit{University College London}\\
London, UK \\
t.jendoubi@ucl.ac.uk}
}

\maketitle

\begin{abstract}
In the context of higher education's evolving dynamics post-COVID-19, this paper assesses the impact of new pedagogical incentives implemented in a first-year undergraduate computing module at University College London. We employ a mixed methods approach, combining learning analytics with qualitative data, to evaluate the effectiveness of these incentives on increasing student engagement. 

A longitudinal overview of resource interactions is mapped through Bayesian network analysis of Moodle activity logs from 204 students. This analysis identifies early resource engagement as a predictive indicator of continued engagement while also suggesting that the new incentives disproportionately benefit highly engaged students. Focus group discussions complement this analysis, providing insights into student perceptions of the pedagogical changes and the module design. These qualitative findings underscore the challenge of sustaining engagement through the new incentives and highlight the importance of communication in blended learning environments.
  
Our paper introduces an interpretable and actionable model for student engagement, which integrates objective, data-driven analysis with students' perspectives. This model provides educators with a tool to evaluate and improve instructional strategies. By demonstrating the effectiveness of our mixed methods approach in capturing the intricacies of student behaviour in digital learning environments, we underscore the model's potential to improve online pedagogical practices across diverse educational settings.
\end{abstract}


\section{Introduction}
In the wake of the COVID-19 pandemic, maintaining active and sustained student engagement has emerged as a paramount challenge in higher education. This challenge has necessitated a critical reassessment of module designs and the development of strategies that foster student engagement to support effective learning. Assessing these design changes requires innovative methods to measure their impact on student engagement and learning accurately. 

Responding to this broader challenge, the Department of Statistical Science at University College London has introduced new pedagogical incentives across its modules. During the pandemic, the first-year computing module, STAT0004: Introduction to Practical Statistics, adopted a flipped classroom model to promote independent learning. Students individually navigated through lecture notes, videos, and quizzes before applying their knowledge in live labs, where they could seek support from the lecturer while practising their coding skills. The module has since been updated with weekly code submissions and weekly feedback, aiming to increase student engagement. Introducing these pedagogical incentives underscores the need to develop an evaluation framework to assess their impact on enhancing student engagement in the module.

A notable gap exists in the effective quantification and modelling of student engagement, which is pivotal when evaluating the impact of pedagogical changes. While student log data extracted from learning management systems, such as Moodle, provide a mechanism to track online student engagement \cite{pardo2014}, \cite{wong&chong2018} caution that reliance on such data alone may overlook critical aspects of the student experience. Recognising that student engagement is multifaceted and challenging to capture comprehensively \cite{sinatraetal2015}, our study adopts a mixed methods approach integrating advanced learning analytics with focus group discussions. The novel application of Bayesian network analysis to Moodle log data enables the dynamic tracking of student interactions with resources over the term, offering a temporal perspective on engagement. This method, enriched by qualitative insights from focus group discussions, presents a holistic model of student engagement. 

Using the STAT0004 computing module as a case study, we showcase the potential of this framework to yield actionable insights, thus equipping educators with the tools to effectively evaluate and inform the adaptation of their teaching strategies to enhance student engagement. The primary objectives of this research are to: 
\textit{(a)} assess the impact of the pedagogical changes in STAT0004;
\textit{(b)} evaluate the overall effectiveness of the STAT0004 module design; and
\textit{(c)} demonstrate a new framework educators can utilise for module evaluations. 

We structure this paper into three main sections: \ref{RW}, which reviews relevant literature; \ref{Meth}, detailing the mixed methods approach of this study; and \ref{R&D}, presenting and interpreting the findings in the context of STAT0004 while considering their implications for educational practice and research in higher education.

\section{Related Works}
\label{RW}
\subsection{Mixed methods in educational research}

Adopting mixed methods in educational research offers a comprehensive approach to understanding complex pedagogical phenomena \cite{nguyen2020}. By integrating quantitative and qualitative data, researchers can capture objective data-driven insights and subjective experiences that often elude quantitative analyses \cite{johnson2007}. Despite its proven effectiveness in evaluating module design interventions \cite{Egilsdottiretal2022} and assessing the influence of pedagogical changes on student engagement \cite{Lin2018}, the adoption of mixed methods in higher education research is surprisingly limited \cite{viberg2018}.

Furthermore, the depth and sophistication of learning analytics tools often fall short in mixed methods studies. Traditional quantitative approaches, such as Likert scale surveys \cite{Muiretal2020}, exam scores \cite{Ahlin2020}, and attendance and login data \cite{Lin2018}, though helpful, only provide surface-level insights. Similarly, qualitative methods, including focus groups \cite{NommeandBirol2014}, structured interviews \cite{Zhiyongandjiaying2022}, and open-ended survey questions \cite{Egilsdottiretal2022}, while insightful for capturing student perceptions, often lack robust quantitative support. This study aims to bridge this gap by combining advanced learning analytics with in-depth qualitative insights. We expect this approach to yield a comprehensive and nuanced understanding of student engagement, paving the way for more effective educational strategies.

\subsection{Bayesian network}

A Bayesian network is a statistical model representing a set of outcomes (nodes in the graph) and their conditional dependencies (arrows in the graph) using a graphical structure informed by observed data. In the context of educational research, these networks can estimate the probability of various student outcomes, such as behaviours and academic performance. When an arrow connects two nodes, it indicates that knowledge of one outcome provides information that can alter the assessed likelihood of the other outcome occurring without implying direct causation. The adaptability of Bayesian networks across different module designs and educational settings \cite{reichenberg2018} highlights their methodological versatility. Furthermore, these networks are visual and accessible, allowing educators of diverse statistical backgrounds to interpret the data.

Bayesian networks have been instrumental in addressing complex educational challenges such as building adaptive tutoring systems \cite{kaser2017}, modelling learning styles \cite{carmona2008}, and predicting student dropout \cite{lacave2018}. Similarly, \cite{xing2021} applied dynamic Bayesian networks to track student interactions in engineering design software. In this study, we employ a dynamic Bayesian network to analyse longitudinal engagement patterns in Moodle log data across instructional resources. This broadens the scope of Bayesian network applications in educational research.

\section{Methodology}
\label{Meth}
\subsection{Context}

This study focuses on the STAT0004 module over the 2022--2023 academic year. While the module spans two terms, totalling 20 weeks, we restrict attention to the first term to expedite evaluating the new changes. This term encompasses ten weeks, with the curriculum covering nine chapters in weeks 1--9 and concluding with a graded assessment in week 10. Live coding labs deliver the module, supplemented by learning resources accessed via Moodle. Most chapters include lecture notes containing content and coding tasks, explanatory videos, a formative quiz, and a code submission point due weekly. Though actively encouraged, participation in the quiz and code submissions is optional, with no direct impact on students' grades. Feedback on the code submissions is verbally provided during the subsequent live lab.

\subsection{Learning analytics}

The objective of the learning analytics component of this study is to quantitatively assess student engagement patterns within the STAT0004 module to evaluate the impact of the new pedagogical incentives. To this end, the analysis leverages a dataset extracted from the Moodle activity logs for STAT0004, detailing the interactions of 204 students with the module's online resources during the first term. This dataset offers a granular view of how students engaged with the module's activities, such as accessing lecture notes, watching videos, completing quizzes, and making submissions. Notably, we cannot capture the impact of the weekly feedback in the log data.

In the dataset, each row records a student's click on the Moodle page for the STAT0004 module, including the time and nature of the interaction. We focused our analysis on synchronous interactions, defined as actions occurring within two weeks of the resource's release date. This was designed to capture regular, day-to-day student interactions with module resources and exclude periods of atypical engagement, such as pre-assessment revision, to align with the study's intentions of evaluating active and sustained student engagement. Synchronous actions were identified using the \texttt{Time} column while the \texttt{Event context} and \texttt{Event name} columns discerned the type of resource interaction. We converted the logs into a binary table, where each entry indicates if a student interacted with a specific resource within the defined synchronous period.

We fitted the Bayesian network using the \texttt{bnlearn} package in R \cite{bnlearnpackage}. The model was constrained to allow stationary or forward connections, meaning the network could only establish relationships between resources of the same chapter or subsequent chapters in line with the chronological order of the curriculum. Lecture notes from chapters 1-3 were excluded from the model as over 95\% of students accessed them, limiting their analytical value. We used bootstrapping to enhance the reliability of the Bayesian network and mitigate the risk of overfitting. This validation process fitted the model 100 times, each with a different subset of the data. We retained connections in at least 50\% of these iterations in the final model, ensuring it included only the most stable and significant relationships between module resources.

The resulting network graphically represents the probabilistic dependencies between resources, offering insights into student behaviours concurrent with the pedagogical changes.

\subsection{Focus group}

We designed the focus group methodology to complement the quantitative data, gathering qualitative insights on students' perceptions of the STAT0004 module and the new pedagogical incentives. All undergraduate students from the Department of Statistical Science were invited via email to participate, encapsulating current and former students of the STAT0004 module. Though initially ten students expressed interest, the focus group comprised five participants, representing a blend of study years, genders, and nationalities, each receiving a £20 compensation for their time. 

We structured the focus group into three sections, each led by a different project team member. The sections sought to: \textit{(1)} gather general feedback on the STAT0004 module; \textit{(2)} collect specific feedback on the newly introduced pedagogical incentives; and \textit{(3)} encourage student input in future module design changes. This structure, while incorporating feedback on the pedagogical changes, also enabled an assessment of current student experiences and promoted student involvement in shaping future educational strategies. Ethical considerations, including informed consent and confidentiality, were rigorously observed.

\section{Results and Discussion}
\label{R&D}
\begin{figure*}[htbp]
\centerline{\includegraphics[width=0.8\textwidth]{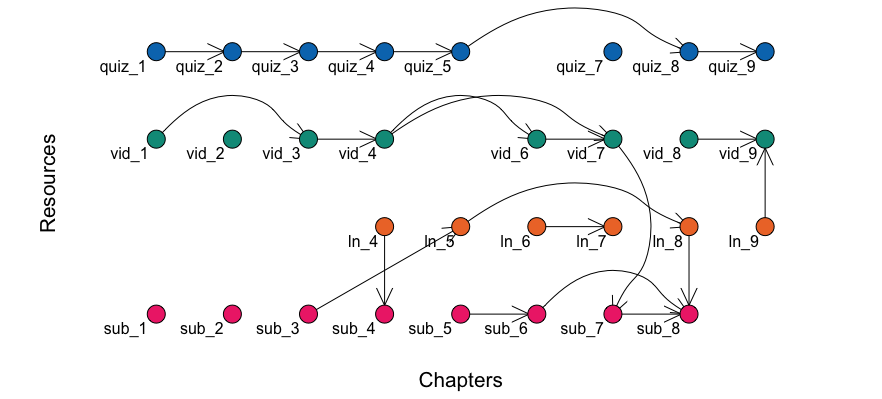}}
\caption{Bayesian Network Depicting Resource Interactions Over Chapters for the STAT0004 Module. This graph illustrates the conditional dependencies between each module chapter's resources (nodes). Nodes are colour--coded by type: blue for quizzes (quiz\_1 to quiz\_9), green for videos (vid\_1 to vid\_9), orange for lecture notes (ln\_4 to ln\_9), and pink for submissions (sub\_1 to sub\_8), where the suffix indicates the chapter number. Directed arrows between nodes represent probabilistic dependencies, where the engagement with one resource may influence the likelihood of engagement with another. Notably, lecture notes from chapters 1 to 3 are absent due to uniform high access rates, while other missing resources were not available.}
\label{fig}
\end{figure*}

Fig. \ref{fig} visualises the Bayesian network analysis of the Moodle activity data. In this graph, each node represents a specific action with a module resource within STAT0004: accessing lecture notes, watching videos, completing quizzes, or submitting code. The arrows between nodes represent conditional dependencies, suggesting how engagement with one resource may relate to engagement with another. These dependencies offer insights into possible relationships rather than direct causation, allowing for an intuitive understanding of patterns in student interactions with the module content. 

We interpret and discuss these results in tandem with the focus group discussions, which offer a qualitative dimension to our understanding of student engagement in STAT0004. 

\subsection{Weekly code submissions}

The Bayesian network graph shows that the nodes representing code submissions for Chapters 1-5, which cover the first half of the term, are disconnected. This observation implies that participation in submissions in any of these early weeks does not necessarily increase the likelihood of submitting work in the following weeks alone. However, this changes after the midpoint, where 80\% of students who submitted code for Chapter 5 also submitted code for Chapter 6, and 85\% of students who submitted code for Chapters 6 and 7 did the same for Chapter 8. Additionally, in the latter half of the term, 83\% of students who watched the video for Chapter 7 also submitted their code.

This insight is supported by the focus group, who found the weekly submissions in STAT0004 initially engaging but noted a decline in participation over time, as these tasks lacked direct implications for grades. Participant A said, ``There's no consequence; you don't need to submit it. It's just more like holding yourself accountable." This observation suggests that while students initially engage with the submissions, their commitment may diminish upon realising their optional nature. To counter this, Participant A suggested that ``...every student cares about their grades..." advocating to integrate submissions into the grading system, with grades awarded for effort, to sustain engagement. 

These findings suggest that the introduction of weekly code submissions has successfully fostered initial active engagement among students. However, these submissions have only maintained the interest of highly engaged students interacting across a broad spectrum of module resources, such as watching weekly videos. This pattern indicates that while these submissions retain the engagement of active participants, they might not have successfully sustained or increased engagement among those displaying lower levels of engagement. This phenomenon could be interpreted as a novel application of the `Matthew Effect' \cite{rigney2010}, aligning with research indicating that educational interventions often benefit already engaged or higher-achieving students more significantly \cite{ottoandkistner2017}. Therefore, it is imperative to implement targeted strategies that captivate and sustain the engagement of all students, especially those at risk of disengaging.

\subsection{Weekly feedback}

While we could not evaluate the impact of the weekly feedback through the Bayesian network analysis, the focus group revealed concerns about the visibility and effectiveness of this new incentive. Participant B highlighted a communication gap: ``To be honest, I'm not aware of any weekly feedback." and suggested making feedback channels more accessible by ``...posting it on Moodle, rather than only discussing it in the lab."

This insight underscores that providing feedback alone is insufficient; it must be effectively communicated and readily accessible to students. The recommendation to employ Moodle for disseminating feedback aligns with \cite{betts2009}, highlighting the critical role of familiar platforms for communication to promote student engagement. Adopting this strategy not only aims to address the current communication gap but also supports the integration of feedback into future quantitative evaluations for a more informed analysis of this pedagogical incentive. 

\subsection{Module design}

The focus group discussions reveal the pivotal role of early and consistent engagement with module resources to remain aligned with the demanding curriculum. Participants emphasised the challenges of independent study and the necessity of self-directed learning strategies and described the module as intensive. Participant C noted, ``This is the kind of module that needs a large amount of time to study by ourselves", while others labelled the module \textit{super hard}. Students identified extensive reading requirements as a barrier to engagement. Participant D shared, ``It takes a long time for me to actually sit down and read the paragraph at the beginning."

The Bayesian network analysis extended this finding by identifying a pattern of sustained engagement: students engaging with the quiz or video resources in one chapter were likely to continue to engage with this same resource in subsequent chapters. For example, 82\% of students who viewed the video for Chapter 3 also accessed the video for Chapter 4. Conversely, students who initially did not engage with a particular resource showed a marked decrease in subsequent participation. Only 14\% of those who skipped the Chapter 1 quiz engaged with the Chapter 2 quiz, and a mere 7\% of students who avoided the Chapter 2 quiz went on to attempt the quiz for Chapter 3.     

These findings underscore the importance of initial engagement with the resources. Extending the findings of \cite{summersetal2020}, which establishes early engagement as a predictor of future behaviour and performance, these insights demonstrate that early engagement with a specific resource might indicate future interaction with that same resource. This observation suggests students may develop routines early, exhibiting reluctance to explore new resources mid-term. Consequently, it becomes imperative to consider strategies that encourage diversified engagement from the onset. This insight might also suggest that interventions to identify and support students showing signs of low initial engagement can be implemented early. 

There were several further interesting observations about the module design. From the Bayesian network analysis, quiz interactions are notably isolated, only linking within themselves. This indicates students' interaction with quizzes was independent of the other resources, which might reflect quiz completion outside the synchronous time frame or some inherently independent nature of quizzes in the module. This aspect warrants further investigation into how different assessment types are perceived and impact student engagement. 

Moreover, participants in the focus group recommended adding live lectures to complement the lab sessions and resources. Participant E compared STAT0004 with a second-year computing module: ``...we have a lecture for that course... I feel like this course is better than [STAT0004]." Others echoed this, indicating a preference for combining theoretical lectures with practical workshops. This feedback highlights an actionable insight for enhancing module design and is a dimension not directly captured in the analytical analysis.

\subsection{Limitations}

A notable limitation of this study lies in the sensitivity of the Bayesian network structure to the specific definition of \textit{synchronous} interactions. The current definition considers interactions within two weeks of release, which was chosen to model regular engagement with module resources. However, even minor adjustments to this definition can result in changes to the structure of the Bayesian network, indicating a lack of robustness to variations in the definition of synchronous interactions. This sensitivity highlights the need for careful consideration in defining engagement periods and suggests that interpretations of the findings must account for this potential variability.

Additionally, despite offering valuable insights, the qualitative aspect of this study is limited by the small number of participants for the focus group. With only five participants, there's a risk the findings may not be representative of the student population. Future research should include a more extensive and varied group of participants or implement more inclusive qualitative methods, such as surveys, to enhance the reliability of the results from the qualitative data.

\section{Conclusion}
\label{Con}

Evaluating and refining pedagogical strategies should be an ongoing process to ensure the effectiveness of educational delivery. This paper introduces a framework that facilitates this, blending advanced learning analytics with qualitative insights to model student engagement. This framework's versatility allows its adaptation to a wide range of educational contexts, resources, and module structures, making it a valuable tool for educators seeking to review and improve their pedagogical approaches using a data-driven and student-centred approach.

Key findings from the STAT0004 module include the necessity of targeted pedagogical incentives for diverse engagement levels and the critical role of effective communication in blended learning environments. Our analysis reveals that early engagement with specific resources is a reliable indicator of future participation, underscoring the importance of early and inclusive engagement strategies.

Building upon the insights gained from the STAT0004 module, our future work will extend the application of this framework to a broader range of modules within the Department of Statistical Science, encompassing both computing and non-computing subjects. This expansion aims to determine the transferability of our findings across different educational contexts and explore how specific aspects of module design influence student engagement.

\section*{Acknowledgments}
We want to thank Ziqing Li and Jingcheng Song for their invaluable assistance as research support students in this project and Prof. Jim Griffin and Prof. Ioanna Manolopoulou for their ongoing supervision and support.

\bibliographystyle{IEEEtran}
\bibliography{ref}

\end{document}